\documentclass[11pt,a4paper]{article}

\usepackage{ascmac}
\usepackage{amsmath}
\usepackage{amssymb}
\usepackage{bm}
\usepackage[footnotesize,bf]{caption}
\usepackage{fullpage}
\usepackage{color}
\usepackage{float}
\usepackage{graphicx}
\usepackage[]{natbib}
\usepackage{subfigure}
\usepackage{txfonts}
\usepackage{threeparttable}
\usepackage{wrapfig}
\usepackage[]{multicol}
\usepackage{wrapfig}
\usepackage{authblk}

\usepackage{floatflt}

\title{\large Complete list of the ASTRO-H Science Working Group}
\date{\vspace{-0.5cm}}
\newcommand{\MakeWhitePaperTitle}{
	\begin{center}
		\begin{figure}
			\vspace{1cm}
			\begin{center}
				\includegraphics[width=0.2\hsize]{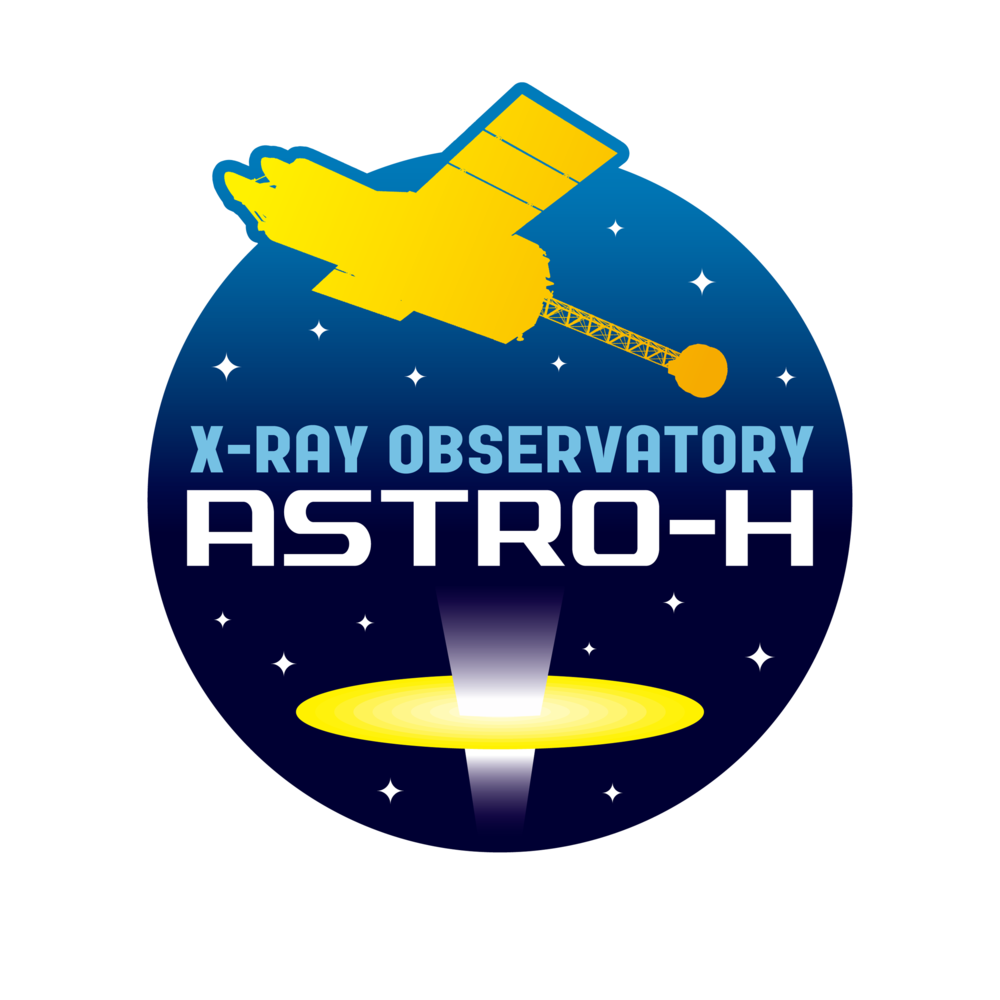}
			\end{center}
		\end{figure}
		\vspace{1cm}
		{\LARGE
		ASTRO-H Space X-ray Observatory\\
		White Paper\\
		}
		\vspace{5mm}
		{\large
		\WhitePaperTitle\\
		}
		\vspace{1cm}
		{
		\WhitePaperAuthors\\
		on behalf of the ASTRO-H Science Working Group
		}
	\end{center}
}

\usepackage{authblk}
\author[a]{Tadayuki~Takahashi}
\author[a]{Kazuhisa~Mitsuda}
\author[b]{Richard~Kelley}
\author[c]{Felix~Aharonian}
\author[d]{Hiroki~Akamatsu}
\author[e]{Fumie~Akimoto}
\author[f]{Steve~Allen}
\author[g]{Naohisa~Anabuki}
\author[b]{Lorella~Angelini}
\author[h]{Keith~Arnaud}
% ...
\affil[a]{Institute of Space and Astronautical Science (ISAS), Japan Aerospace Exploration Agency (JAXA), Kanagawa 252-5210, Japan}
\affil[b]{NASA/Goddard Space Flight Center, MD 20771, USA}
\affil[c]{Astronomy and Astrophysics Section, Dublin Institute for Advanced Studies, Dublin 2, Ireland}
\affil[d]{SRON Netherlands Institute for Space Research, Utrecht, The Netherlands}
\affil[e]{Department of Physics, Nagoya University, Aichi 338-8570, Japan}
\affil[f]{Kavli Institute for Particle Astrophysics and Cosmology, Stanford University, CA 94305, USA}
\affil[g]{Department of Earth and Space Science, Osaka University, Osaka 560-0043, Japan}
\affil[h]{Department of Astronomy, University of Maryland, MD 20742, USA}
% ...

\begin{document}

\newcommand{\WhitePaperTitle}{Plasma Diagnostic and Dynamics of the Galactic Center Region}
\newcommand{\WhitePaperAuthors}{
	K.~Koyama~(Kyoto~University),
	J.~Kataoka~(Waseda~University),
	M.~Nobukawa~(Kyoto~University),
	H.~Uchiyama~(Shizuoka~University),
	S.~Nakashima~(JAXA),
	F.~Aharonian~(Max~Planck~Institute~f\"ur~Kernphysik~\&~Dublin~Institute~for~Advanced~Studies),
	M.~Chernyakova~(Dublin~Institute~for~Advanced~Studies),
	Y.~Ichinohe~(JAXA~\&~University~of~Tokyo),
	K.~K.~Nobukawa~(Kyoto~University),
	Y.~Maeda~(JAXA),
	H.~Matsumoto~(Nagoya~University),
	H.~Murakami~(Tohoku~Gakuin~University),
	C.~Ricci~(Kyoto~University),
	L.~Stawarz~(JAXA),
	T.~Tanaka~(Kyoto~University),
	T.~G.~Tsuru~(Kyoto~University),
	S.~Watanabe~(JAXA),
	S.~Yamauchi~(Nara~Women's~University), and
	T.~Yuasa~(RIKEN)
}
\MakeWhitePaperTitle

\begin{abstract}
The most characteristic high-energy phenomena in the Galactic center (GC) region is the presence of strong K-shell emission lines from  highly ionized Si, S, Ar, Ca, Fe and Ni, which form the Galactic Center X-ray Emission (GCXE). These multiple lines  suggest that the GCXE is composed of at least two plasmas with temperatures of $\sim$1 and$\sim$7~keV. The GCXE also exhibits the K-shell lines from neutral Si, S, Ar, Ca, Fe and Ni atoms. A debatable issue is the origin of the GCXE plasma; whether it is a diffuse plasma or integrated emission of many unresolved point sources such as cataclysmic variables and active binaries. Detailed spectroscopy for these lines may provide a reliable picture of the GCXE plasma. The origin of the K-shell lines from neutral atoms is most likely the fluorescence by X-rays from a putative past flare of Sgr A$^*$. Therefore {\it ASTRO-H} may provide unprecedented data for the past light curve of Sgr A$^*$.  
All these lines may provide key information for the dynamics of the GCXE, using possible Doppler shift and/or line broadening. This paper overviews these line features and the previous interpretation of their origin. We propose extended or revised science with the {\it ASTRO-H} observations of some select objects in the GC region. 
\end{abstract}

\maketitle
\clearpage

\tableofcontents
\clearpage

\section{Background and Previous Studies}
\subsection{GC Hot Plasma}
Diffuse (unresolved) X-ray emission from our Galaxy was discovered just 7 years after the birth of X-ray astronomy \citep{Co69}. The {\it HEAO-A2} data were systematically searched, and revealed  diffuse X-rays in the 2--10 keV band along the  Galactic plane extending  by about 7~kpc with half thickness of 241$\pm$2~ pc, and hence called as the Galactic ``Ridge'' X-ray emission (GRXE) \citep{Wo83}. The {\it EXOSAT} data also show an extended emission over more than $100^{\circ}$ along the Galactic plane but only $\sim 2^{\circ}$ across it \citep{Wa85}. 

The K-shell line from He-like iron at 6.7~keV was first discovered with the {\it Tenma} satellite  \citep{Ko86a}. The {\it Ginga} satellite found that the 6.7~keV line profile shows a sharp peak ($|l|<1^{\circ}$ and $|b|<0^{\circ}.5$) around the Galactic center (GC) region \citep{Ko89}. 
This peak emission is separately named as the Galactic Center X-ray Emission (GCXE). The origin of the GRXE and the GCXE has been a mystery since its discovery.

\begin{figure}[bph] 
\begin{center}
\includegraphics[width=1.0\linewidth]{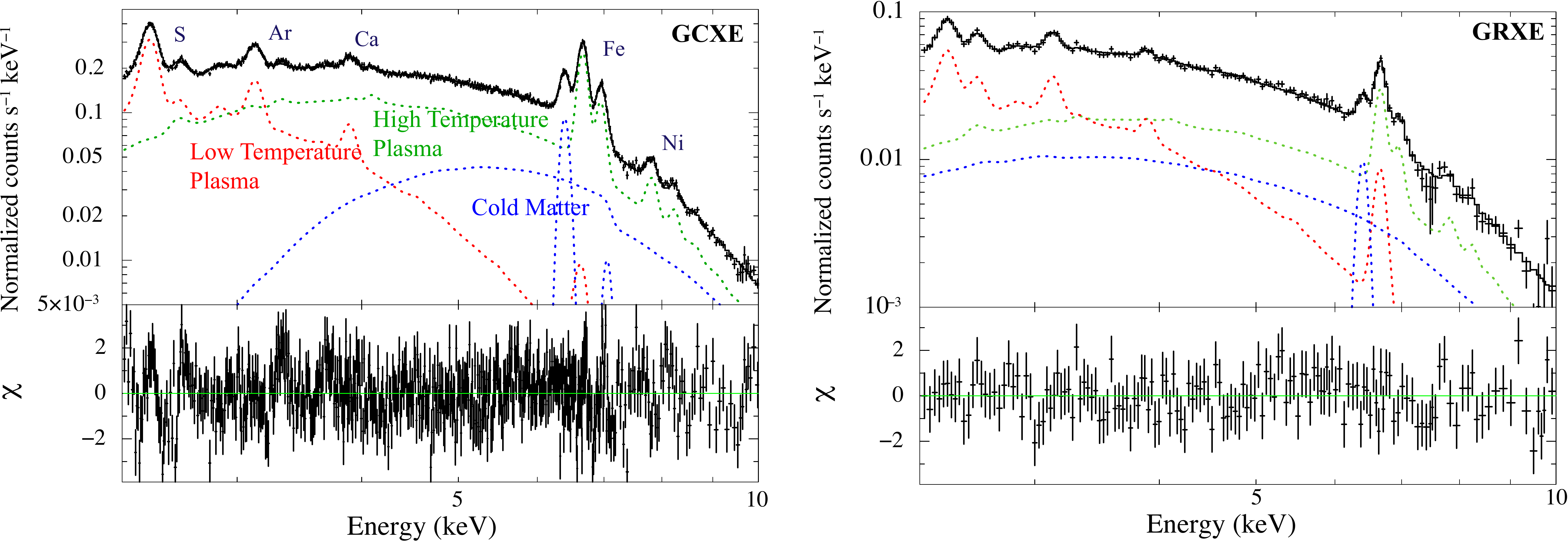}
\end{center}
\caption{X-ray spectra of the GCXE (left) and GRXE (right) observed with {\it Suzaku}/XIS \citep{Uc13}.  Many K-shell lines from highly ionized atoms, S, Ar, Ca, Fe and Ni are found. The dotted lines are the best-fit  two-temperature ($\sim 1$ and $\sim10$~keV) thermal plasmas and cold matter (power-law  +  Fe I K lines) components.} 
\label{fig:GC_Sp}
\end{figure}

Recently, the superior spectral capability of the {\it Suzaku} satellite provide excellent X-ray spectra of the GCXE and GRXE as are  shown in figure \ref{fig:GC_Sp}. The co-existence of the highly ionized S and Fe lines indicates that the GCXE and GRXE are due to hot plasmas consisting of at least two temperature components: low temperature plasma (LP; $kT\sim 1$~keV) and high temperature plasma (HP; $kT\sim 7$~keV). 
If the plasmas are truly ``diffuse'' gases,  the temperatures are too high to be confined by the gravity of the Galaxy, and hence the plasma gas should escape from the Galactic center and plane regions.  In order to sustain such hot plasmas in our Galaxy, huge amounts of energy need to be supplied continuously. 
This energy can be a result of high energy activities in the Galaxy [e.g. jets of a supermassive black hole (SMBH) Sgr A$^*$ \citep{Ko96}, reconnections of the Galactic magnetic field \citep{Ta99}, presence of a large amount of supernovae remnants \citep{Ko86b}]. 

\begin{figure}[bpt] 
\begin{center}
\includegraphics[width=1.0\linewidth]{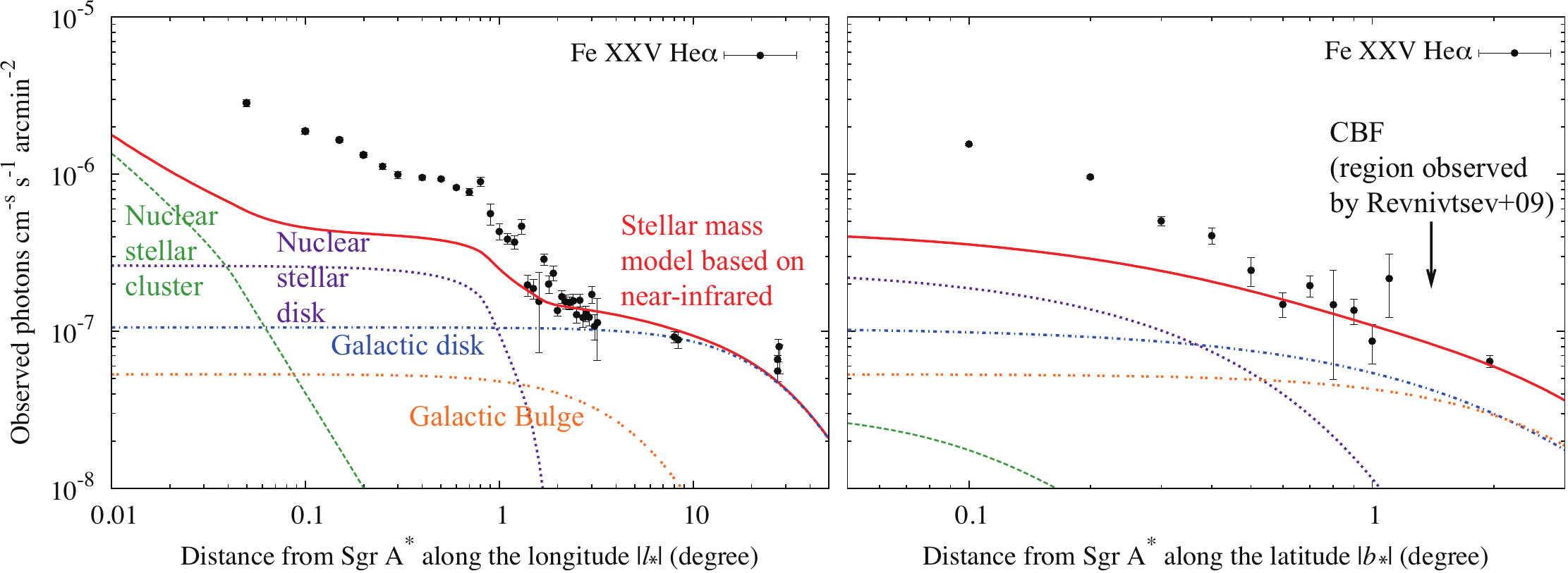}
\end{center}
\caption{Fe~XXV~He$\alpha$ line  profile along the Galactic longitude (left, $b\sim0^{\circ}$) and latitude (right, $l\sim0^{\circ}$).  The green and purple lines are  the stellar  mass distribution in the   central nuclear stellar cluster and nuclear stellar disk. The blue and orange lines represent the Galactic disk (ridge) and bulge. The sum is presented by the  red solid line. These intensities are  normalized to the disk data \citep{Uc11}.} 
\label{fig:GCXEvsIR}
\end{figure}

To solve these potential problems, a ``point source'' scenario for the origin of the GCXE and GRXE has been also proposed.  \citet{Wo83}, for example,  pointed out that integrated emission of many faint X-ray stars like cataclysmic variables (CVs) and active binaries (ABs)  contribute up $43\pm18$\% of the GRXE. 
The long-exposure ($\sim$1 Ms) observation of {\it Chandra} resolved about 90\% of the 6.7~keV line from the {\it Chandra} Bulge Field (CBF) at $(l,b)=(0^{\circ}.1,-1^{\circ}.4)$ into many faint point sources \citep{Re09}. 
\citet{Re09}, \citet{Ho12} and \citet{Mo13} argued that these point sources in the CBF are mainly CVs  and ABs.
The CBF corresponds to the Galactic disk and bulge components and not the GC (nuclear stellar bulge and disk) components (see figure~\ref{fig:GCXEvsIR}). 
Thus hereafter we regard the CBF as a part of the GRXE region.  
The {\it Suzaku} spectra of the GRXE in the $2-50$~keV band can be explained as an assembly of CVs \citep{Yu12}. 
The extrapolated number density of these unresolved point sources can be enough to explain the luminosity of the GRXE \citep{Sa06}.
These results may support an idea that the major component of the GRXE is the superposition of many X-ray point sources, possibly combination of CVs and ABs.  
This scenario however contains a serious problem, the equivalent widths (EWs) of Fe~XXV~He$\alpha$ and Fe~XXVI~Ly$\alpha$\footnote{We use the terminologies of He$\alpha$ and He$\beta$ for the transition to K-shell (quantum number $n=1$) from higher excited states (quantum number $n=2$ and 3) of He-like atoms, respectively.}
 in the GRXE are significantly larger than those of any mixture of known CVs and ABs \citep{Uc13,Wa14}. 

To answer this problem, unresolved faint CVs and ABs should have about 2--3 times larger iron line EWs than those of the known bright CVs and ABs.  In the faint sources, either the iron abundance is larger than known sources or other unknown mechanisms to enhance the iron line are at work. 

\begin{figure}[htbp] 
\begin{center}
\includegraphics[width=1.0\linewidth]{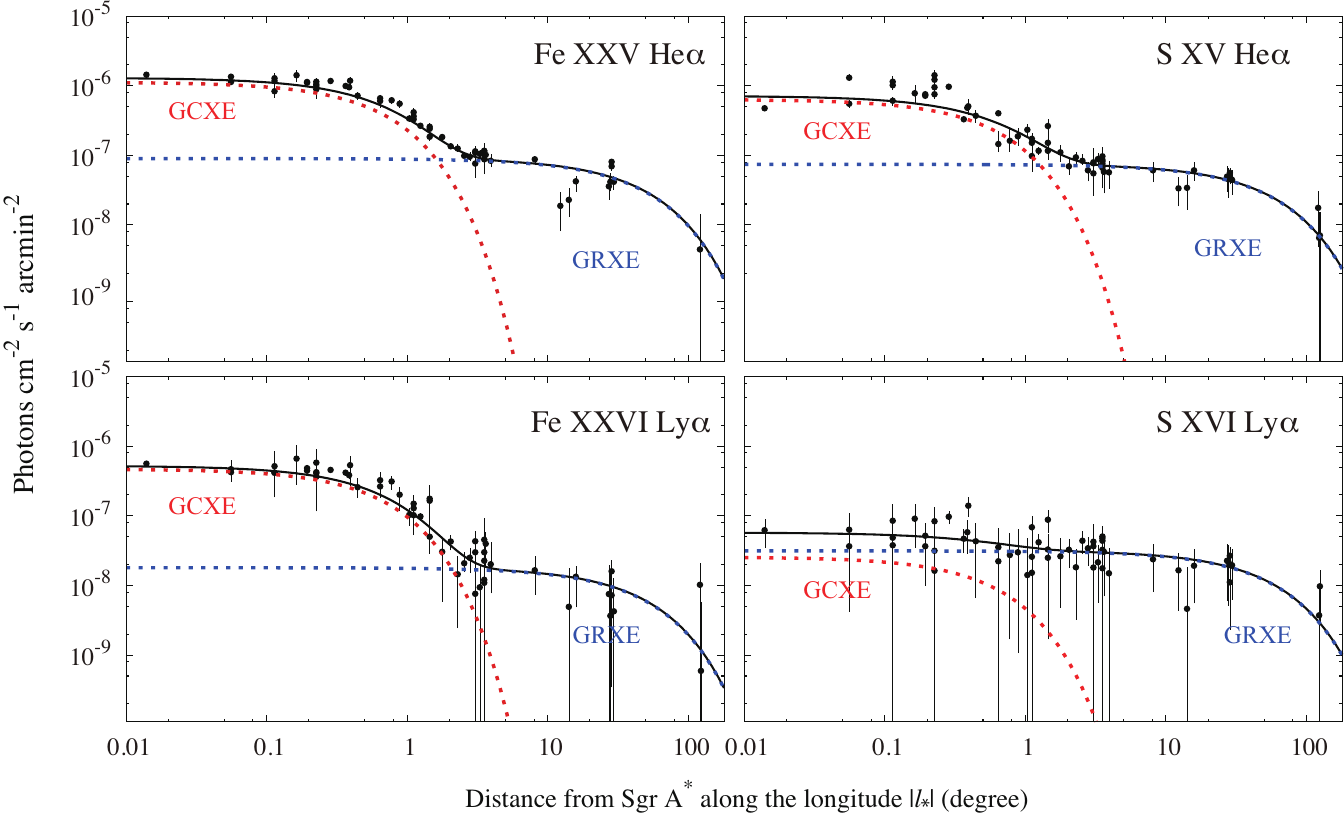}
\end{center}
\caption{Flux distributions of Fe~XXV~He$\alpha$ (upper left), Fe~XXVI~Ly$\alpha$ (lower left), S~XV~He$\alpha$ (upper right) and S~XVI~Ly$\alpha$ (lower right) \citep{Uc13}. 
Red and blue dotted lines show the GCXE and GRXE components, respectively.} 
\label{fig:GCXE_GRXE_Dis}
\end{figure}

Figure \ref{fig:GCXE_GRXE_Dis} is the flux distributions of Fe~XXV~He$\alpha$, Fe~XXVI~Ly$\alpha$, S~XV~He$\alpha$ and S~XVI~Ly$\alpha$ along the Galactic plane. The longitudinal distribution profiles are different among these lines; the Fe~XXVI~Ly$\alpha$ flux at the GCXE ($|l|<1^{\circ}$) shows larger excess relative to those of the GRXE compared to Fe~XXV~He$\alpha$, and vice versa for S~XV~He$\alpha$ and S~XVI~Ly$\alpha$. 
Thus the temperatures of the GCXE and the GRXE are different such that the temperature of LP in the GCXE is lower than the GRXE and HP at the GCXE is higher than the GRXE. 

Figure \ref{fig:GCXEvsIR} shows the comparisons between profiles of the Fe~XXV~He$\alpha$ line, which is a typical component of the HP, and the stellar mass distribution based on near-infrared observations \citep{Uc11}. 
The emissivity of Fe~XXV~He$\alpha$ per stellar mass in the GCXE is about two times larger than that in the GRXE. \citet{He13} also found similar excesses with {\it XMM-Newton}. 
The authors mainly used the {\it COBE} results \citep{La02} so that the stellar mass profile has a large uncertainty \citep{He13}. 
\citet{Ni13}, on the other hand, used high-quality SIRIUS/IRSF data for the near-infrared observation and confirmed the Fe~XXV~He$\alpha$ line excess at the GCXE above the prediction from the stellar number distribution in the GRXE.

These results strongly suggest that the origin of the GCXE is different from that of the GRXE, whose origin is likely to be superposition of faint X-ray stars. 
In fact, unlike in the case of the GRXE, only 10--40\% of the GCXE could be resolved into point sources even with long-exposure {\it Chandra} observations ($\sim600$--900 ks; \citealt{Mu03, Re07}).  
If major fraction of the GCXE is point sources origin for the GCXE, more serious problem of the iron line EWs than that in the GRXE appears, because the enhancement of the iron EWs above the known point sources is much larger in the GCXE than in the GRXE. 
Thus we argue that truly diffuse plasma is a major contributor to the GCXE.
The plasma may be related to past high energy activities (flare of Sgr A$^*$) in the GC (see  the following sections), responsible for the X-ray echo (section \ref{sec:x-echo}) and jet-like structures (section \ref{sec:GCflare}).
The putative large  flares from Sgr A$^*$  might produce hot plasma, as is found around the micro-quasar SS433 \citep{Ko94}. 

Another possibility is superposition of supernova remnants (SNRs).  In fact,  {\it Suzaku} has found many new SNR candidates in the GC region. Some of these SNRs have high-temperature plasmas which are not found in the other regions of the Galaxy (e.g. G0.61$+$0.01; \citealt{Ko07SgrB}). In any case, the GCXE should contain key information about the extreme conditions in the GC region. The very high-resolution and wide-band spectra of the GCXE by SXS and HXI will be a clue to reveal the origin of these extremes.

\subsection{X-ray Echo: K-shell Lines from Neutral Atoms}\label{sec:x-echo}
\begin{figure}[hbpt] 
\begin{center}
\includegraphics[width=.8\linewidth]{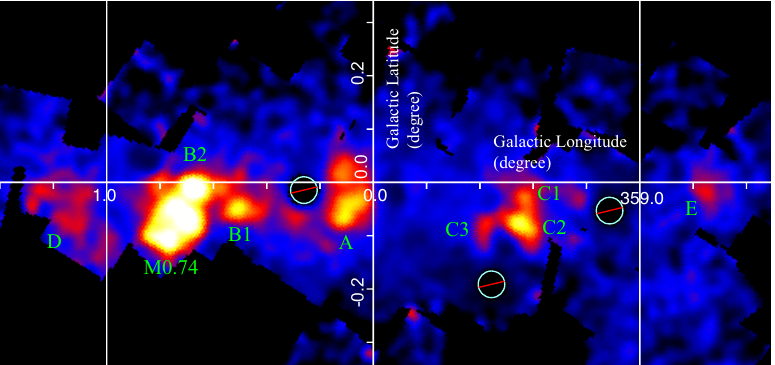}
\end{center}
\caption{The XRNe complex  near the Galactic center: Sgr~D, B, A, C, and E (from the left to the right). This image is a 6.4~keV EW map obtained with {\it Suzaku}.} 
\label{fig:EWmap}
\end{figure}

The other K-shell lines notable in the GCXE and GRXE are those of  neutral atoms, in particular the 6.4~keV line from neutral iron (figure \ref{fig:EWmap}).
The spectra also exhibit a strong Fe K-shell absorption edge. 
These spectral features support the X-ray Reflection Nebula (XRN) scenario (e.g. \citealt{Ko96}) in which X-rays are due to fluorescence and Compton scattering by the cold gas of the giant molecular clouds (MCs) in the central molecular zone (CMZ).
Since the fluxes of the XRNe require a high X-ray luminosity for an irradiating source, which is brighter than any of the brightest Galactic X-ray sources, the potential candidate of the illuminating X-ray source should be the SMBH, Sgr A$^*$. 
 
\begin{figure}[hbp] 
\begin{center}
\includegraphics[width=.6\linewidth]{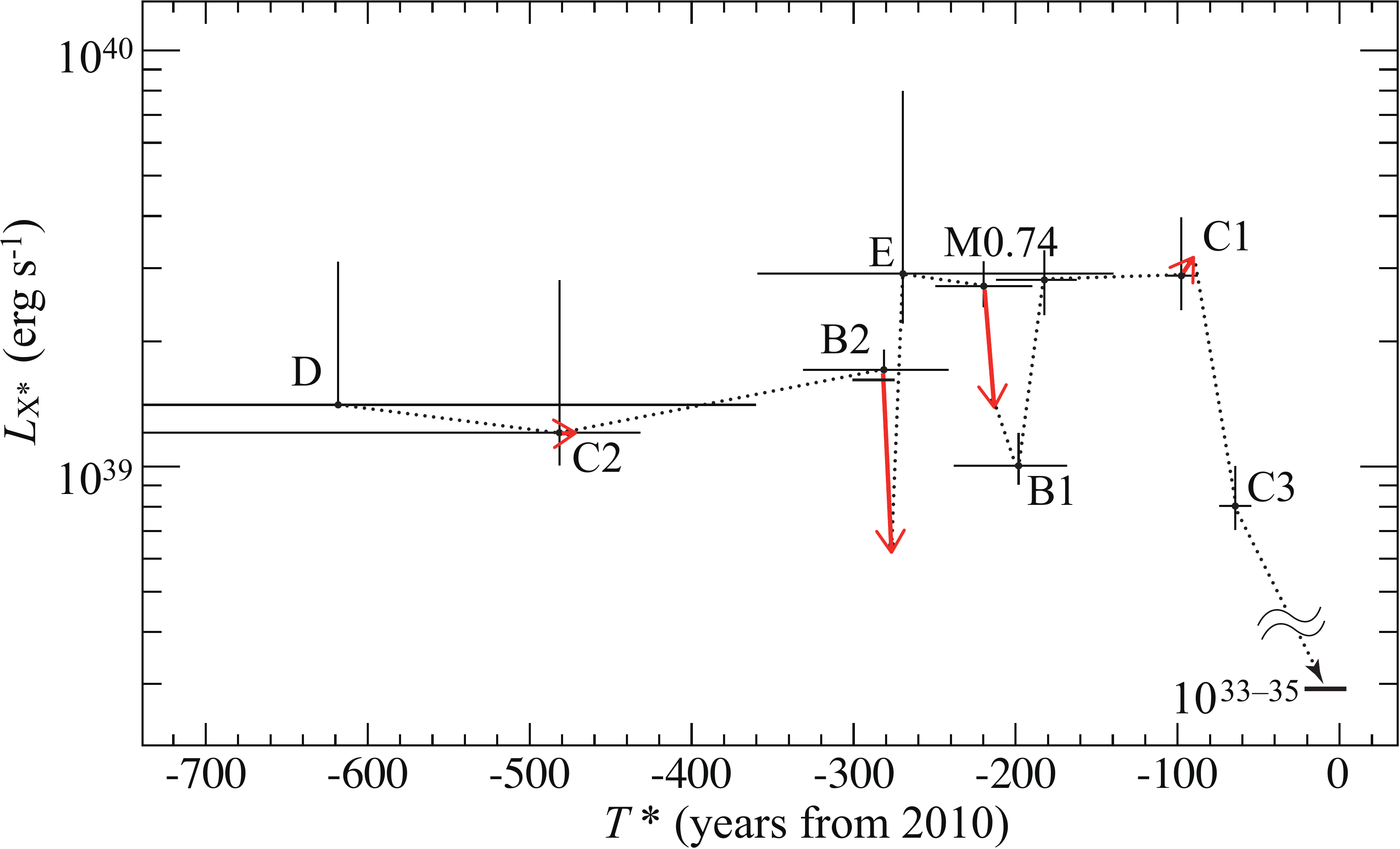}
\end{center}
\caption{Light curve of Sgr A$^*$ in the past $\sim$ 100--700 years as derived 
by the X-ray echo scenario \citep{Ry13}. 
Red arrows indicate short-term flux increase or decrease. 
A label of each data point corresponds to the XRN shown in figure~\ref{fig:EWmap}
} 
\label{fig:SgrA*LC}
\end{figure}

The observed spectrum of the GCXE is a mixture of the GC plasma and XRNe. Since X-rays of the GC plasma located behind the XRNe are largely absorbed, while those coming from the plasma in front the XRNe are not, the combined spectra have positional information of XRNe in the GC plasma.  \cite{Ry09,Ry13} analyzed the GCXE spectra as the sum of emission coming from the GC plasma in front and behind the XRNe, and additional emission of XRN itself. The results give line-of-sight positions of the XRNe in the GC plasma. Combining with the projected position (2-dimension), we construct the three-dimensional distribution of the XRNe. Then  using the flux of XRNe, we can obtain the fluxes of Sgr A$^*$ at different epochs (figure \ref{fig:SgrA*LC}).  The X-ray luminosity of Sgr A$^*$ was about $10^6$ times higher than it is now. Then the flux dropped down to the current level about 100 years ago.

\begin{figure}[hbpt] 
\begin{center}
\includegraphics[width=.46\linewidth]{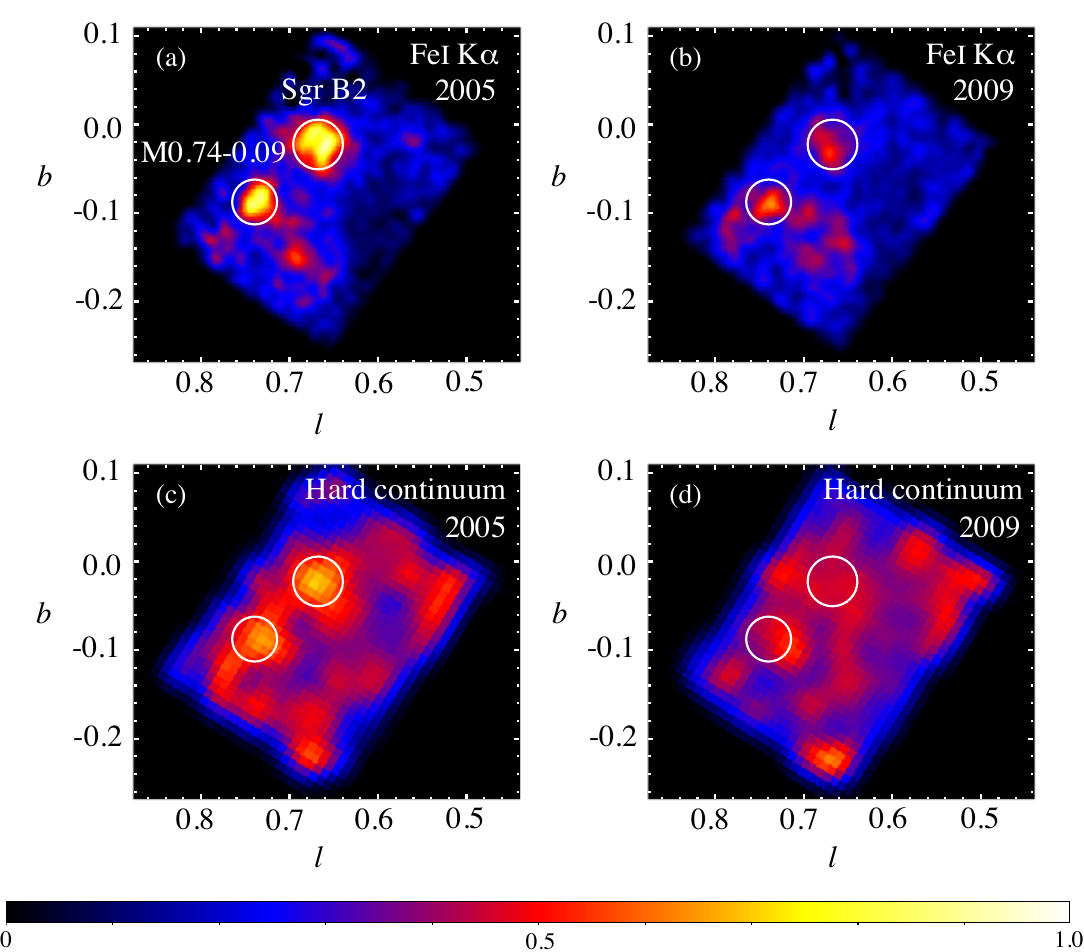}
\includegraphics[width=.52\linewidth]{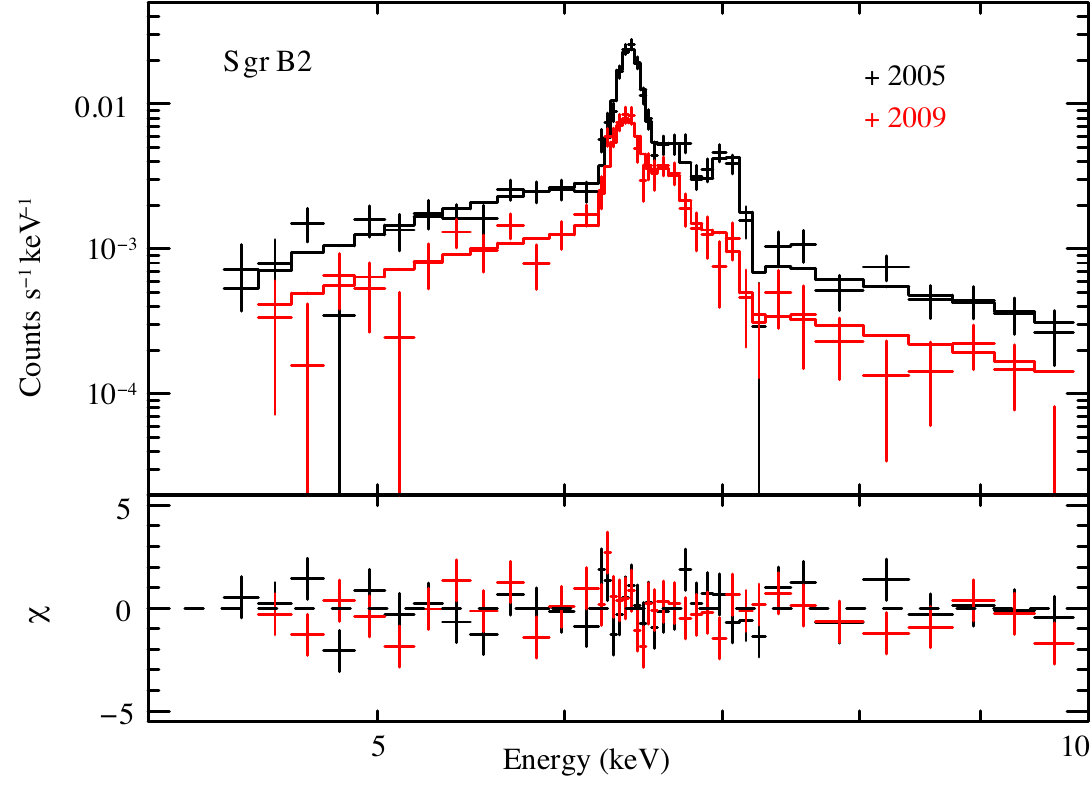}
\end{center}
\caption{Left: The images of  short-term time-variability of the Sgr B2 cloud from 2005 to 2009. Right:  X-ray spectra of Sgr B2 \citep{No11}.} 
\label{fig:SgrB2}
\end{figure}
 
Some of the XRNe exhibit time variability on a timescale of few years (figure \ref{fig:SgrB2}), (e.g. \citealt{Ko08, In09, Te10, Po10}). Thus the high flux level between 100--700 years ago cannot be a single flat-top flare, but would be due to multiple 
short-time flares of the peak flux about (1--$3)\times10^{39}$ erg s$^{-1}$. 
\cite{Po10} argued that apparent superluminal motion of a light front illuminating a molecular nebula might be due to a source outside the MC (such as Sgr A$^*$ or a bright and long outburst of an X-ray binary). Monte Carlo simulations by \cite{Od11} presented the
detailed morphologies and spectra of the reflected X-ray emission for several realistic models of Sgr~B2.

Other than in the giant molecular clouds, the weaker 6.4~keV emission is prevailing all over the GC region. 
The origin has still been an open issue. \cite{Ch12} suggests that the extended 6.4~keV emission can be 
explained by ambient molecular gas reflecting the past X-rays from Sgr A$^*$ like the XRNe. An alternative explanation is due to fluorescence and 
bremsstrahlung or inverse bremsstrahlung emission of low energy electrons or ions (e.g. \citealt{Yu02,Do09}). 
We do not have crucial information, such as line broadening and equivalent width of the 6.4~keV line,
to constrain the origin, which will be obtained by {\it ASTRO-H}.

\subsection{Other Relics of Big Flares of Sgr A$^*$ }\label{sec:GCflare}

\begin{figure}[bt] 
\begin{center}
\includegraphics[width=.45\linewidth]{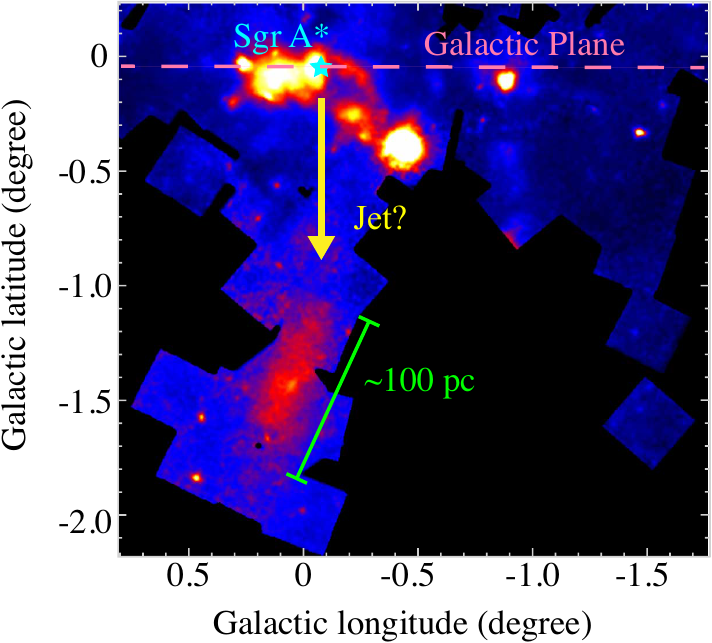}
\includegraphics[width=.45\linewidth]{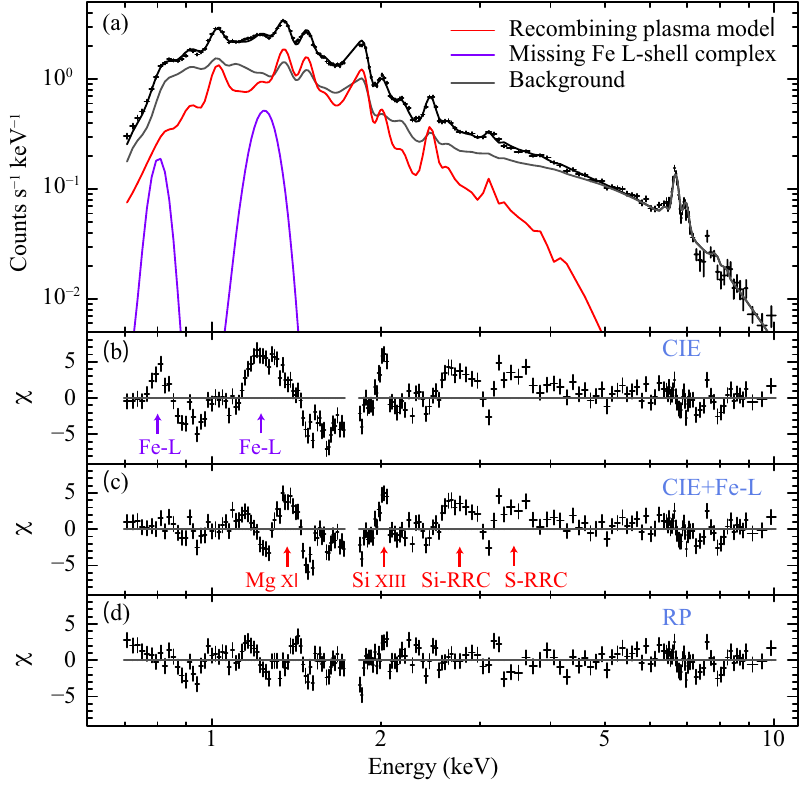}
\end{center}
\caption{Left: Jet-like plasma at the Galactic south (Galactic bulge jet). 
Right: The X-ray  spectrum and model fits. The CIE fit shows clear excess of at the Si~XV~Ly$\alpha$ line and the RRC structures of Si and S at around 2--4~keV (the 2-nd and 3-rd panels),  while these excesses are disappeared with the RP fit (1-st and 4-th panels) \citep{Na13}. } 
\label{fig:GCJet}
\end{figure}

{\it Suzaku} recently discovered a thermal emission of about 0.5~keV at around ($l$, $b$)= (0$^\circ$, $-1^\circ$.5) (see figure \ref{fig:GCJet} left; \citealt{Na13}).  
Remarkable features  of this plasma is that it has a jet-like  structure ejected from Sgr A$^*$ (here, the Galactic bulge jet). Also the X-ray spectra cannot fit with a collisional ionization equilibrium (CIE) plasma model, leaving clear excesses at the Si~XV~Ly$\alpha$ line and the radiative recombination continuum (RRC) structure of  Si and S at around 2--4~keV. The spectra are nicely fitted with a recombining plasma (RP) model as is shown in figure \ref{fig:GCJet} (right). 
One plausible scenario is that the almost fully ionized  (at least, for  Si and S) plasma was made by a jet-like activity (flare)  of Sgr A$^*$  $\sim2\times10^5$ years ago, and is now in a recombining phase.  

The {\it Fermi} satellite discovered the large scale GeV gamma-ray emissions called  ``the {\it Fermi} bubble''. The {\it Fermi} bubbles are extending  about $50^{\circ}$  (or $\sim10$~kpc) above and below the GC \citep{Do10,Su10}, suggesting that  a starburst or nuclear outburst happened near/within the GC  $\sim10^{6-7}$ years ago. This scenario is consistent with the origin of the North Polar Spur (NPS) proposed by \citet{So00}. The {\it Fermi} bubble is  spatially correlated with the ``haze'' observed with {\it WMAP}, and the edge follows along  NPS in the {\it ROSAT} X-ray maps. The {\it Suzaku} X-ray spectra show $kT\sim 0.3$~keV plasma with large  $N_{\rm H}$, consistent with a scenario of shock-heated Galactic halo during  the bubble expansion \citep{Ka13}.

In addition to the  recent flares (100--700 years ago), Sgr A$^*$ likely experienced an energetic flare 
in the far past of $\sim2\times10^5$ years ago,  and  even more energetic flares $\sim 10^{6-7}$ years ago. 
We thus propose that energetic flares of Sgr A$^*$ may not be a rare event in the past. 
The high and low temperature plasmas of $kT\sim7$~keV and $kT\sim1$~keV, as well
as a lower temperature plasma of $kT \sim 0.3$--$0.5$~keV in the {\it Fermi} bubble
and the Galactic south jet \citep{Na13} may be produced in a single (or multiple) flare events in GC, tracing the history of ``shock evolution'' from the GC to the outermost regions, the {\it Fermi} bubble. 

A {\it Chandra} 1~Ms observation of Sgr A$^*$ revealed many small flares ($L_{\rm X}=10^{34-36}$~erg~s$^{-1}$) at a frequency of once per day \citep{Ne13}.
From the past activity history, we suggest that big flares with $L_{\rm X}>10^{39}$~erg~s$^{-1}$ has occurred once per thousand years, and possibly $L_{\rm X}>10^{41-43}$~erg~s$^{-1}$ flares  once per million years.
A radio observation has detected that a gas cloud with a few Earth masses is falling into Sgr A$^*$, and hence we may have a chance to observe a big flare in the near future \citep{Gi13}. 
Since Sgr A$^*$ is the nearest SMBH, {\it ASTRO-H} should not miss this one-in-thousand year's chance.  

\subsection{Serendipitous Discovery; Sterile Neutrino }\label{sec:Seren}

\begin{figure}[hptb] 
\begin{center}
\includegraphics[width=1.0\linewidth]{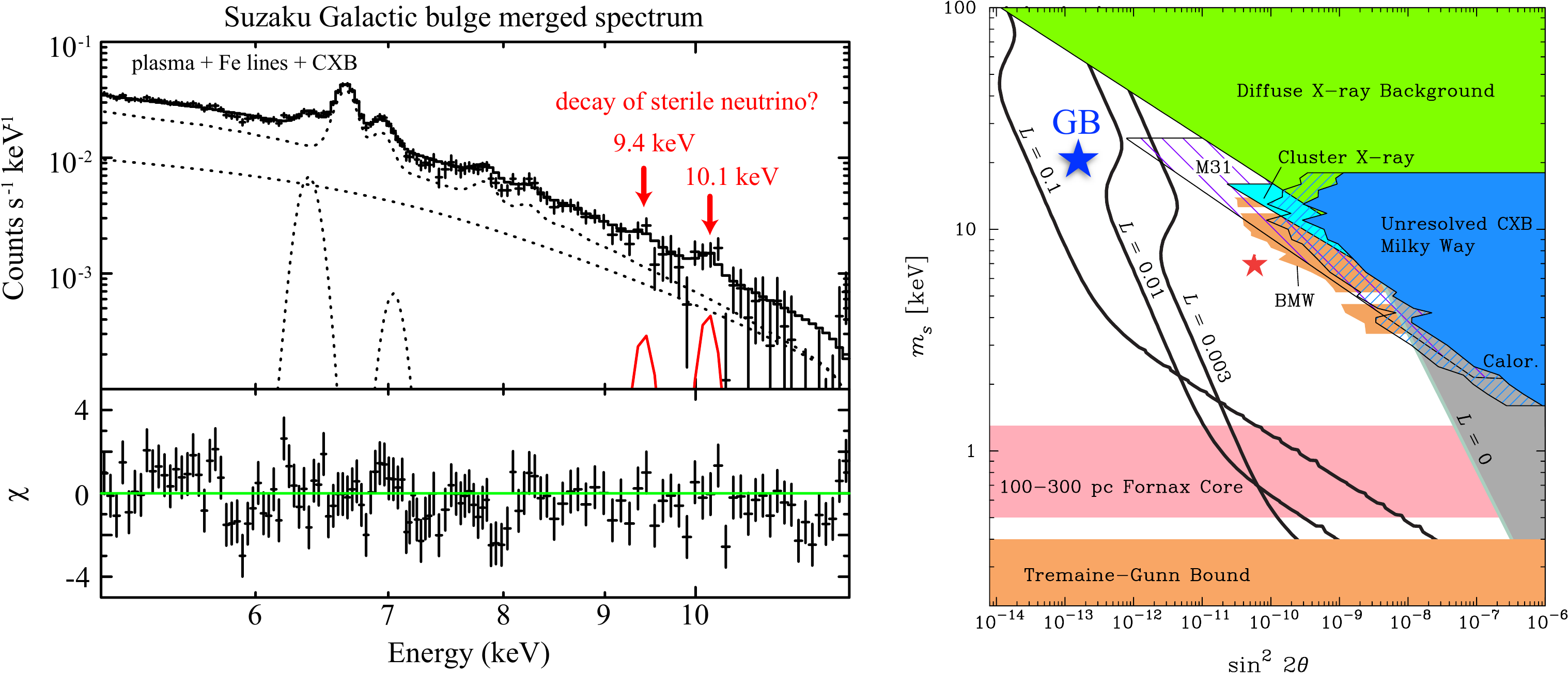}
\end{center}
\caption{
Left: {\it Suzaku} spectrum obtained from the Galactic bulge region with {\it Suzaku}. 
Fitting with the same model as GCXE and GRXE spectra shown in figure~\ref{fig:GC_Sp} (HP, Fe I K$\alpha$/K$\beta$, and CXB) 
gives line-like residuals at the energies of no atomic line, 9.4 and 10.1~keV. 
Right: the plot on the mass-mixing angle plane of our ``sterile neutrino'' (blue star), 
which is located below the current upper-limit region. Original figure is taken from \citet{Bu14}.
} 
\label{fig:Neutrino}
\end{figure}

The Galactic center region is revealed to be a ``treasure box'' of high energy astronomy.
In fact, we have made many serendipitous discoveries in addition to the results given in the previous subsections.  We show one example, a hint of the sterile neutrino line.  
In the {\it Suzaku} GCXE spectrum, possible sterile neutrino signal is found at 8.7~keV \citep{Pr10, Ch11}.  However the GCXE is very strong X-ray background,
and hence possible artificial structure due to the strong X-rays may not be excluded.  
The Galactic Bulge (GB) region has a lower background, and we find  more faint line-like features at 9.4~keV and 10.1~keV with the {\it Suzaku} GB observations (figure \ref{fig:Neutrino}). Neither atomic line nor instrumental structure exists at these energies (the closest line is Au-L at 9.7~keV). 
Although, at this moment, the significance level of the line emission is only 2.5$~\sigma$, these residuals should not be ignored, because it can be a hint of a new important discovery, may or may not be a decay line of sterile neutrinos.  
In fact many outstanding discoveries in the GC, particularly the 6.4~keV echoes (and hence past Sgr~A$^*$ flares), are established by an initial faint signature in the {\it ASCA} and {\it Chandra} observations \citep{Ko96, Mu01}. 
We note that, similarly, there is a new suggestion of a weak line at $\sim3.5$~keV in the X-ray spectra of clusters of galaxies with deep exposures observations of {\it XMM-Newton} satellite \citep{Bu14}.

\section{Prospect of {\it ASTRO-H}}
The general objectives in the Galactic center science have been overviewed in section 1.  The immediate  objectives of {\it ASTRO-H} is to give more secure constraint on the origin of  the K-shell lines from highly ionized and neutral atoms with the best  quality  spectra. 
The  high resolution spectra will provide unprecedented information on plasma diagnostics, and also a good tool for understanding the dynamics of the Galactic cool gas (XRNe) and Galactic hot plasma (mainly GCXE). 
In the following subsections, we will separately discuss for the regions of Radio Arc  (\ref{sec:RadioArc}), Sgr A West (\ref{sec:SgrAWest}), 
Sgr A East (\ref{sec:SgrAEast}), Sgr A$^*$ (\ref{sec:SgrA*}) and Sgr B/C/D/E (\ref{sec:SgrBCDE}).

\subsection{Radio Arc}\label{sec:RadioArc}
\begin{figure}[tb]  
\begin{center}
\includegraphics[width=.6\linewidth]{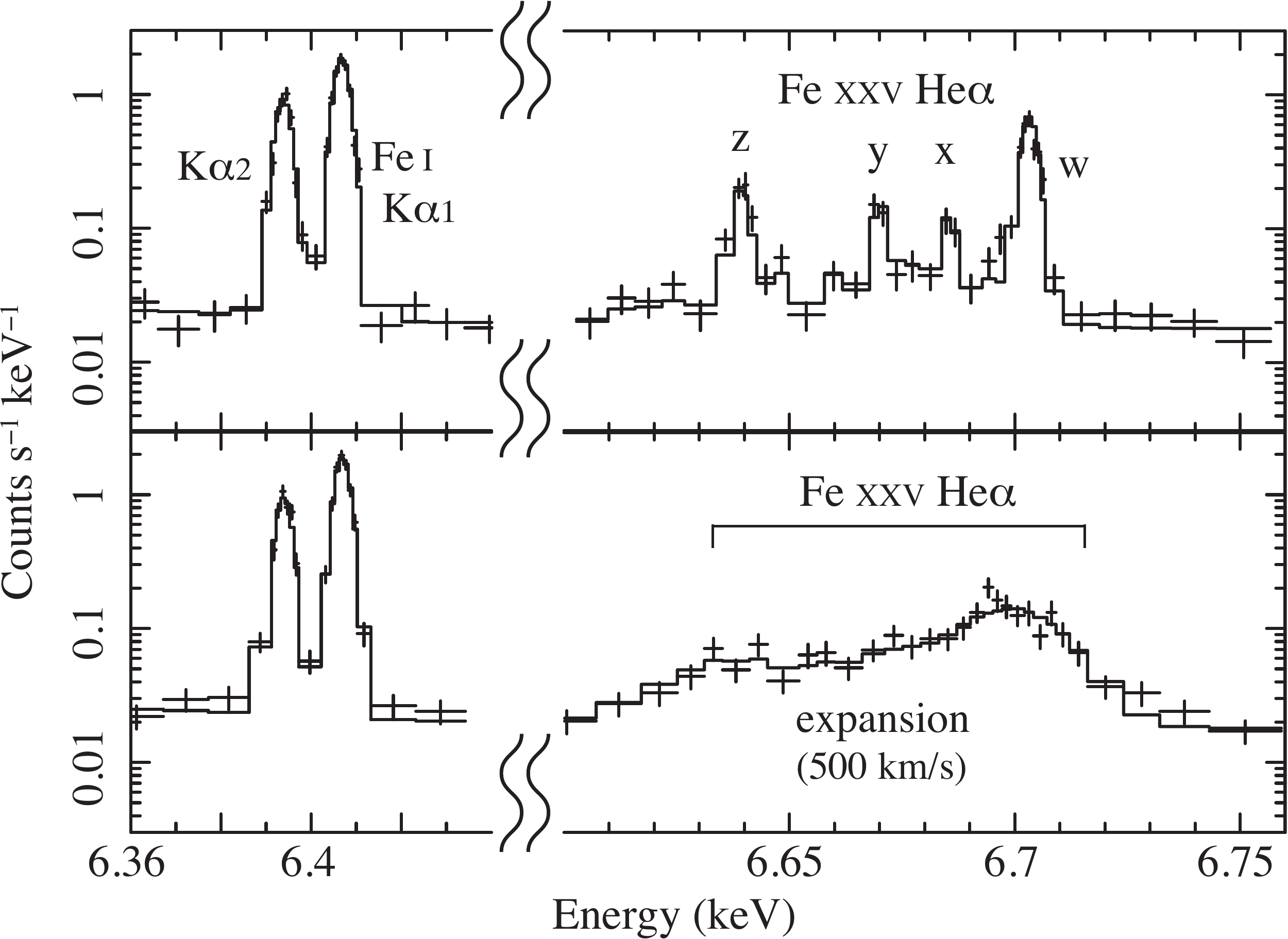}
\end{center}
\caption{SXS spectra of the Radio Arc region with a 100~ks exposure. Upper panel: The 6.4~keV and 6.7~keV (for the case of point source origin) line. Lower panel: Same as the upper panel, but the 6.7 keV line is expanding with the velocity of 500~km~s$^{-1}$ (diffuse origin)}
\label{fig:RAsim}
\end{figure}

This is the brightest region of the GC plasma and the XRNe, and hence this region would 
be the best position to study the nature of both the  GC hot plasmas and  cool XRNe. As is 
shown in the simulated SXS spectrum given in figure \ref{fig:RAsim}, these two major 
components are clearly separated. 
The 6.4~keV lines come from cold molecular clouds (XRNe), hence have no thermal 
broadening. If the 6.7~keV lines are of diffuse origin in a free expanding plasma, the 
velocity is $\sim3000$~km~s$^{-1}$. However, {\it Suzaku} measured the line broadening 
(FWHM) of 37$-$41~eV (F~XXV~He$\alpha$), 29$-$35~eV (Fe~I~K$\alpha$) and 
0$-$23~eV (Fe~XXVI~Ly$\alpha$) \citep{Ko07}. 
The line width of Fe XXV He$\alpha$ is biased by $\sim$25~eV (1~$\sigma$) due to the 
mixture of resonance $(w)$, inter-combination $(x, y)$ and forbidden lines $(z)$ within the 
energy band of 60~eV. Thus, we can safely assume that the intrinsic line broadening is 
$\sim$10~eV or  $\sim$500 km s$^{-1}$ (1~$\sigma$) for the expansion velocity.
This smaller velocity than that of the free expansion would be  due to some confinement 
mechanisms of the hot plasma (e.g. magnetic field).  If we determine the expansion 
velocity, then we can constrain the plasma confining mechanisms. 
We show the simulated SXS spectra with 100~ks exposures in figure~\ref{fig:RAsim} with 
the expansion velocity 0~km~s$^{-1}$ (Upper) and 500~km~s$^{-1}$ (Lower).
We constrain the expansion velocity of the 6.7~keV line to be $500\pm80 $~km~s$^{-1}$ 
(hereafter, errors are at 90\% confidence levels).

If the 6.7~keV lines are of point source origin, although less likely as we discuss in section~1, the width of the 6.7~keV line should be narrow, similar to the 6.4~keV line. Thus by the width measurement of the 6.7~keV line, we can constrain the origin.

\begin{figure}[hbpt]  
\begin{center}
\includegraphics[width=.6\linewidth]{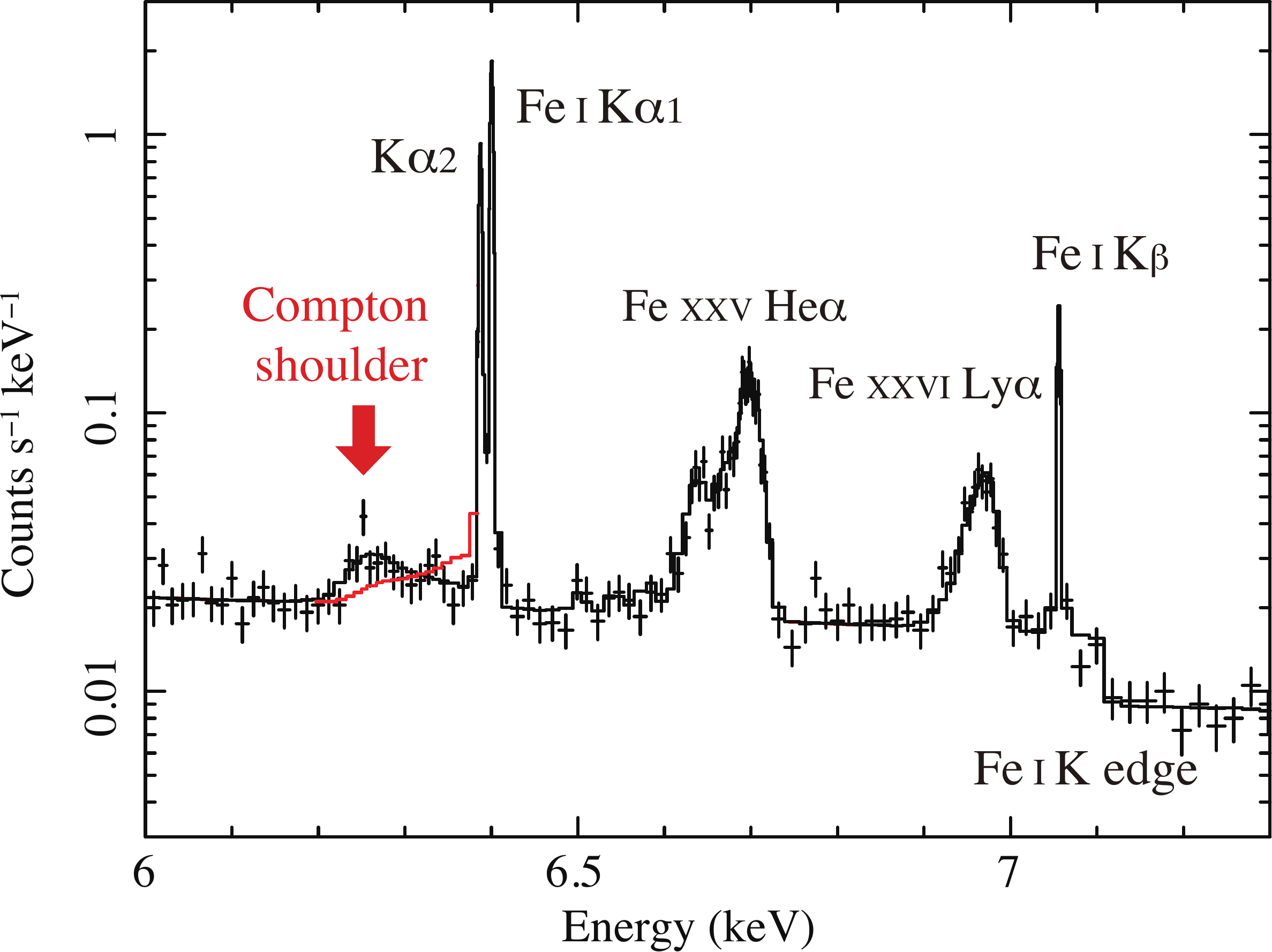}
\end{center}
\caption{Same as the lower panel in figure~\ref{fig:RAsim} but in 6.0--7.3~keV band. 
Data and black lines are of backside reflection ($\theta=180^\circ$). 
Red line is of normal reflection ($\theta=90^\circ$)}
\label{fig:Shoulder}
\end{figure}

Since this region is the bright in the 6.4~keV line, we can easily measure the Compton shoulder of the 6.4~keV line with a reasonable exposure time.  The lowest energy of the Compton shoulder gives the scattered angle ($\theta$); $\theta$ is the angle  between two lines of Sgr A$^*$--Radio Arc and our line-of-sight.  Thus it provides the line-of-sight position of the XRNe. 
The flux of the Compton shoulder is proportional to the hydrogen column density $N_{\rm H}$ of the scattered gas. 
We can also detect Fe K-edge absorption, which immediately tells us the $N_{\rm Fe}$ of these clouds \citep{No11}.
The equivalent width (EW) of the 6.4~keV line depends on $\theta$, because Compton scattering flux (continuum band) is proportional to 
$1+{\rm cos}^2(\theta)$. Thus if we can determine the Fe abundance, the EW 
can also be an indicator of the scattering angle $(\theta)$.  
We simulate the Compton shoulder profile for the backside reflection (scattering angle $\theta=180^\circ$) and normal reflection ($\theta=90^\circ$). With a 200~ks exposure, we can determine $\theta$ with an error of $\sim 10^\circ$, corresponding to the time-bin of 50 year in the look-back light-curve of Sgr A$^*$. This error is a similar value  of \citet{Ry13}, and hence the light curve can be smoothly combined.  
Together with the scattering angle $\theta$, $N_{\rm Fe}$ and the 6.4~keV line flux, we can accurately derive the period and luminosity of the past activity of 
Sgr A$^*$.

\subsection{West of Sgr A$^*$ }\label{sec:SgrAWest}
This is a mirror position of Radio Arc with respect to Sgr A$^*$.  Unlike the Radio Arc region, this region is weak in the 6.4~keV line, hence rather pure GC hot plasma spectra will be obtained. 
The He$\alpha$ line complex and Ly$\alpha$  would be essentially the same as the Radio Arc region (see fig~\ref{fig:RAsim}).
The major objective is to resolve the fine structure of the He$\alpha$ line complex. The line ratios of resonance/forbidden and inter combination lines give good plasma  diagnostics. We note that pointing positions discussed in sections~\ref{sec:RadioArc} and \ref{sec:SgrAWest} are very close to those of the {\it Suzaku} pointing described in \cite{Ko07}.

\subsection{Sgr A East}\label{sec:SgrAEast}
This is a young mix-morphology (MM) SNR, whose plasma temperature is the highest ($kT \sim 3-6$~keV) among the Galactic SNRs.  
Within the central $\sim$5~pc ($\leq 2'$) from Sgr A$^*$, Sgr A East is a unique object surely identified as an SNR. It may provide a rare probe to search the central 5~pc region.
The {\it Suzaku} spectrum shows clear emission of Fe~XXV~K-shell line and a hint of a Cr line \citep{Ko07SgrAEast}.   
Unlike many young-intermediate aged  SNRs, the spectrum is not in IP, but rather in CIE \citep{Ko07SgrAEast}.  
However we cannot exclude the more exotic case of RP, because many of MM SNRs that were previously  known as a CIE plasma are now found to be RP (e.g. \citealt{Yamaguchi09,Oz09,Oh11, Uc12, Sa12, Ya13, Oh14, Ya14}). 

The shell size is 8$\times6~$pc$^2$ with the density of $\sim1$~cm$^{-3}$ \citep{Ko07SgrAEast}. 
The expansion velocity is constrained by the {\it Suzaku} observation; The line width of Fe 
XXV He$\alpha$ and Fe XXVI Ly$\alpha$ are 41$-$47~eV and 0$-$35~eV, respectively. 
The line width of Fe~XXV~He$\alpha$ is biased by $\sim$25~eV (1$\sigma$) (see section 
2.1). Thus the intrinsic line broadenings is $\sim$10~eV or 
$\sim$500~km~s$^{-1}$ $(1\sigma)$ of the expansion velocity.
Then  the dynamical age and ionization parameter $nt$ are $\sim$6000~ years and $\sim2\times10^{11}$~cm$^{-3}$~s, respectively.  Thus if normal SNR, the spectrum 
should be IP with the Fe XXV He$\alpha$ line energy of $\leq$6.6~keV. However 
{\it Suzaku} gives the energy to be 6.7~keV \citep{Ko07SgrAEast}, 
the same as the RP of W49B.  

\begin{figure}[bt]  
\begin{center}
\includegraphics[width=.45\linewidth]{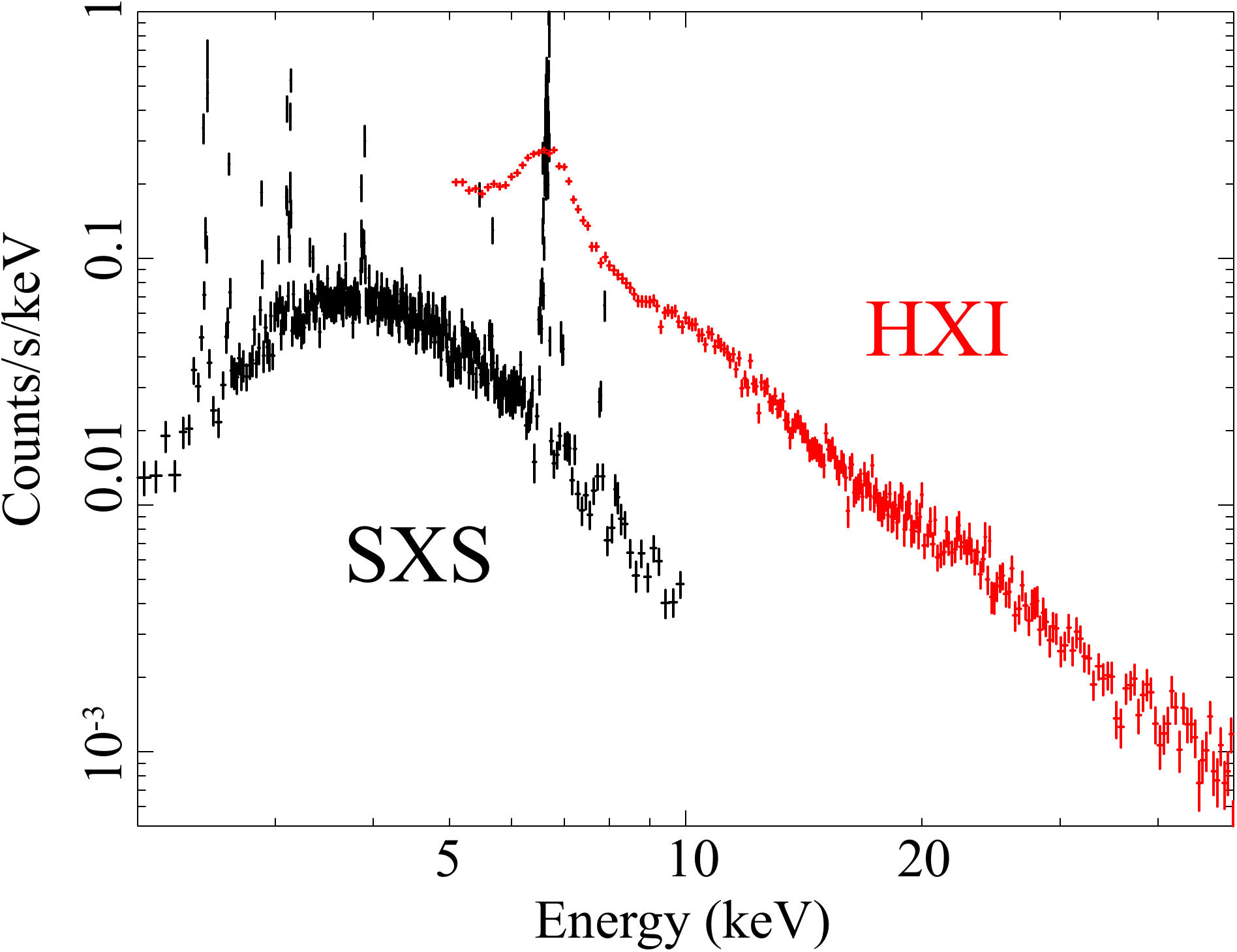}
\includegraphics[width=.46\linewidth]{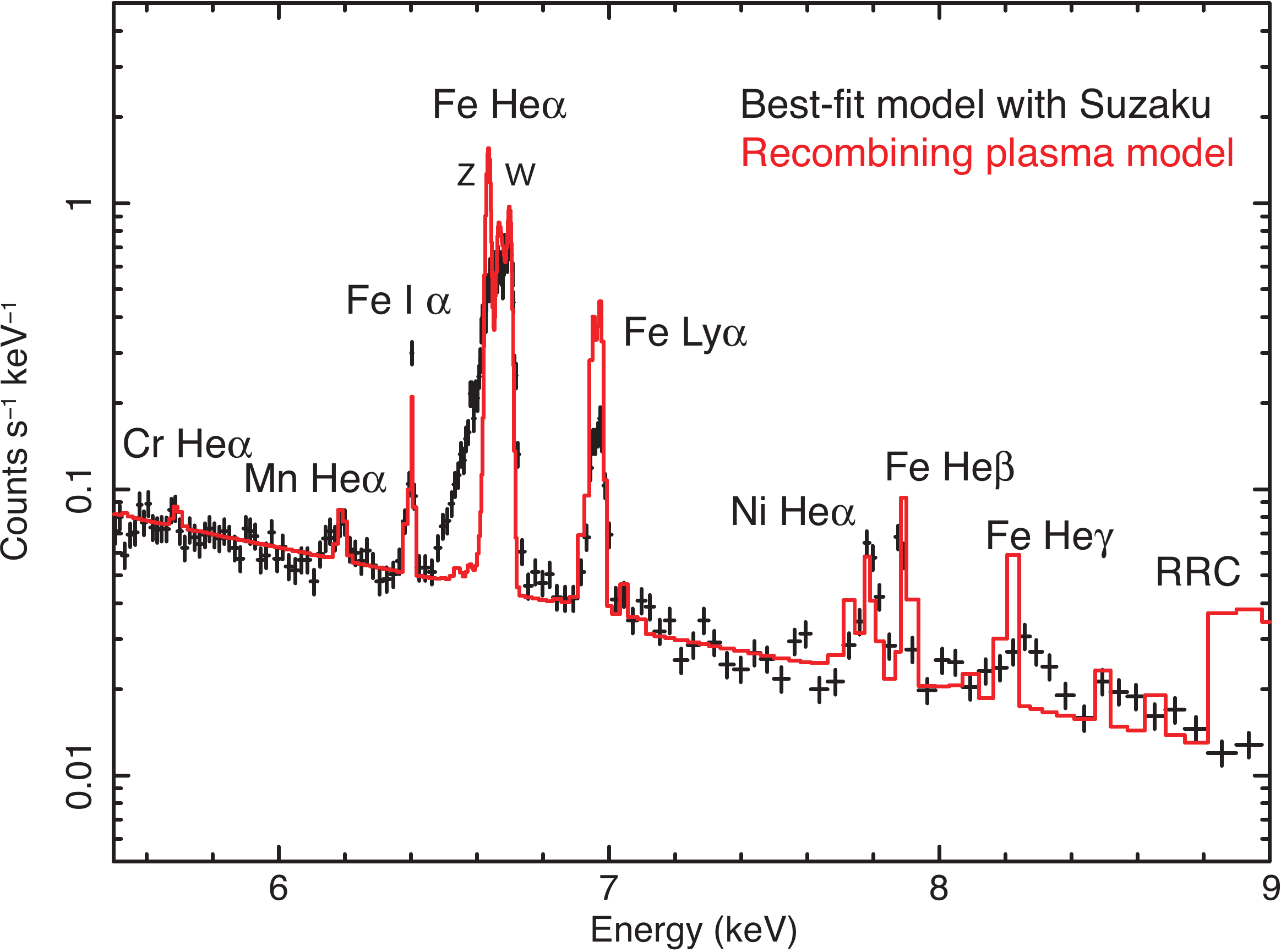}
\end{center}
\caption{Left: Simulation of wide band spectrum (SXS+HXI) for Sgr A East with a 100 ks exposure. Right: Simulation of SXS in the energy band from the Fe~XXV~He$\alpha$ complex to 9~keV. The black data and line are  CIE model, while the red line is a RP model of $nt=2\times10^{11}$ s~cm$^{-3}$. The resonance and forbidden lines are indicated by $w$ and $z$. In CIE, $w\geq z$, while in RP, $w \leq z$.  Thermal expansion of 500~km~s$^{-1}$ is assumed}
\label{fig:SgrAEastsim}
\end{figure}

We can obtain SXS spectra up to 10~keV, and check whether a radiative recombination continuum (RRC) structure of Fe at about 9~keV is present or not.  
Another critical test to distinguish between CIE (or IP) and RP would be done with the help of Fe~XXV~He$\alpha$ fine structure lines. In CIE (IP),  the line flux of $w$ is larger than $z$, while in RP $w$ is smaller than $z$ (see figure $\ref{fig:SgrAEastsim}$ right). 

Sgr A East is in the unique circum-Sgr A* medium: dense star populations, strong magnetic field, high-energy particle field (seen by the GeV/TeV $\gamma$-rays), strong shearing by the Galactic rotation (see figure \ref{fig:GCDynamics} Left) and so on.  These  environments may make the SNR to be unusual, which will be revealed by RP, hard X-ray tail (HXI) or else. Figure $\ref{fig:SgrAEastsim}$ right shows the simulated RP spectrum in $nt=2\times10^{11}$~cm$^{-3}$~s on the faked CIE data (100~ks exposure). 
In RP, we separate $z$ and $w$ line with $z/w$ excess over the CIE value with 20~$\sigma$ level. Together with the discovery of the RRC edge at 8.8~keV, we obtain new and detailed structures of the Sgr A East plasma. Then we investigate a unique evolutional history of Sgr A East in the GC environment. 

The age is rather old, and hence the elongated morphology along the Galactic plane can be due to the shear by the large differential rotation near at the center (see section~\ref{sec:GCDynamics} and figure \ref{fig:GCDynamics}). 
 Then the Doppler shift of the Fe-K lines will follow the stellar rotation in figure \ref{fig:GCDynamics} ($\sim +100$~km~s$^{-1}$), 
since the hot plasma emitting Fe-K lines is relaxed with the Galactic rotation.  
On the other hand, if the SNR is young, the plasma could have not enough time for the relaxation and the motion of the plasma may not follow the Galactic rotation curve. 
The spectrum of Sgr A East may be contaminated by the GCXE, but with the help of the Sgr A West observation discussed in section~\ref{sec:SgrAWest}, we can determine the contribution of the GCXE very accurately.

Hard-X/gamma-rays from a compact region around Sgr A$^*$ were discovered with {\it INTEGRAL} 
in the energy range between 20 and 100~keV \citep{Be04}, and also with HESS between 0.16 and 10~TeV \citep{Ah04}. These sources may be  slightly offset from Sgr A$^*$, and likely to be associated to Sgr A East. However, given the 13$^{'}$ resolution of {\it INTEGRAL}/IBIS, whether these emissions are associated to Sgr A East or not was unclear. Also  exact nature of highly energetic emission from the Sgr A complex is still unknown.  The wide-band spectroscopy of {\it ASTRO-H} between 0.4 and 80~keV with much better angular resolution of $1^{'}.7$ will reveal the exact nature of the hard X-ray emission (figure \ref{fig:SgrAEastsim}) for the first time. 
 Using the information of the Fe~XXV He$\alpha$ fine structure, RRC, possible shearing effects and hard X-ray tail structure, we may constrain the physical parameters on this unique SNR. 

\subsection{Sgr A$^*$}\label{sec:SgrA*}
This is a super massive black hole (SMBH) in our Galaxy. Sgr A$^*$ is in the same SXS field of the Sgr A East SNR, the position is only about $1'$ off-set from the center of the SNR.  
Unlike many SMBHs, the X-ray flux is extremely lower than the Eddington limit at present. 
However, as we argued in section~\ref{sec:GCflare}, energetic flares of Sgr A$^*$ may not be a very rare event, possibly one per 1000 years.  

In the {\it ASTRO-H} era, we may be lucky to witness such an event \citep{Gi13}. 
We should not miss the one-in-thousand year's chance. 
We can get  the  wide band spectrum (SXS, SXI, HXI and SGD) of the flare state of Sgr A$^*$, for the first time.  
In the previous studies, we were only able to constrain the Sgr A$^*$ spectrum in the past flare to be a simple power-law with $\Gamma \sim 2$, due to the limited information of the XRN spectra.
The wide band spectra of {\it ASTRO-H} will give new information: a Fe K line from the SMBH, Sgr~A$^*$ itself and a cutoff shape at the hard X-ray band.

By comparing with the past flare investigated in sections \ref{sec:x-echo}--\ref{sec:GCflare}, we can study whether the past flare was  attributed to the Radiation Inefficient Accretion Flow (RIAF), or to a standard accretion flow and whether phase transition of accretion occurred between the past flares and  low (quiescent or weak flare) state at present.

The observed 511~keV line emission most recently confirmed by {\it INTEGRAL}  (e.g., \citealt{Je06,We08}) toward the GC can be explained naturally in a standard framework of RIAF, if the typical accretion rate was about 1000 times higher than the current rate in the past 10~Myrs \citep{To06}. 
The outflow energy by such an accretion rate is expected to be $10^{56}$~erg (or $3\times 10^{41}$~erg~s$^{-1}$).
Thus identification of the origin of 511~keV line emission using the SGD, together with unresolved connection to  GeV/TeV gamma-rays recently reported by the {\it Fermi}-LAT and HESS (e.g., \citealt{Chernyakova11}) are key science left in the {\it ASTRO-H} era.

If we concentrate on the Fe~XXV~He$\alpha$ line in the GCXE from the close vicinity of Sgr A$^*$, we may discover a relic of the past activity of Sgr A$^*$ of $\sim 10^{41-42}$~erg~s$^{-1}$ at $\sim2\times10^5$ years ago \citep{Na13}. 
The putative flare would photo-ionize (e.g. Fe XXV $\longrightarrow$ Fe XXVI, in the GCXE) to make recombining plasma (RP). Using the density $(n)$ of 0.1~cm$^{-3}$ \citep{Uc13}, the recombination parameter is $nt\sim5\times10^{11}$ cm$^{-3}$~s, hence the plasma is likely to be still in RP phase. 

\begin{figure}[hbtp] 
\begin{center}
\includegraphics[width=.6\linewidth]{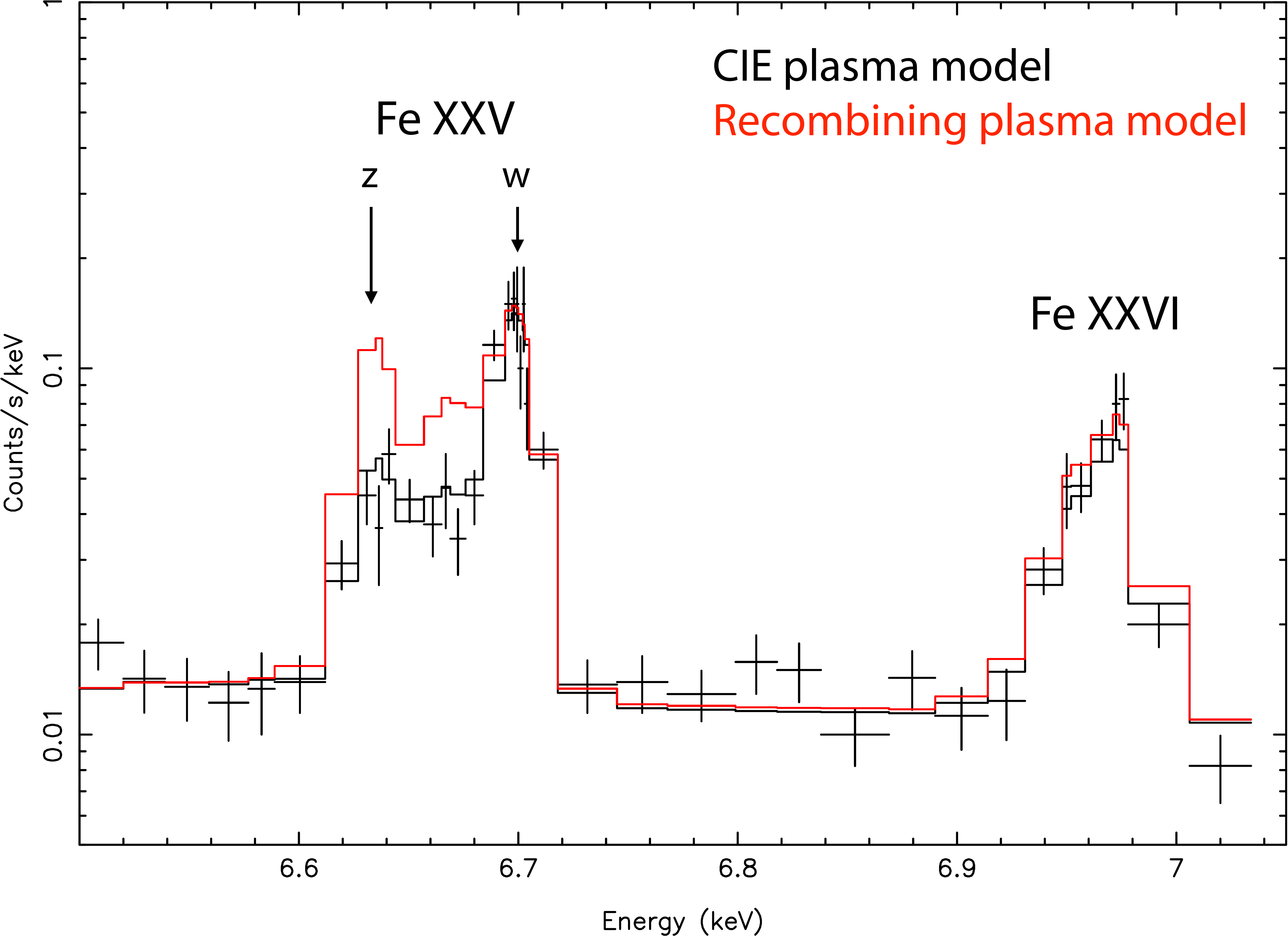}
\end{center}
\caption{A 100 ks simulation of the GCXE in the close vicinity of Sgr A$^*$ with SXS. 
Thermal expansion of 500~km~s$^{-1}$ is assumed.}
\label{fig:SgrA-star}
\end{figure}

Using the flux ratio of resonance $(w)$ and forbidden $(z)$ lines in the Fe XXV He$\alpha$, we investigate whether GCXE plasma near Sgr A$^*$ is RP or not. This study was impossible with the XIS resolution.   A 100~ks simulation of SXS gives $z/w$ to be $0.99\pm0.11$ in RP of $nt\sim 5\times10^{11}$ cm$^{-3}$~s, while collisional ionization equilibrium (CIE) is 0.43 (figure~\ref{fig:SgrA-star}). Thus RP will be  established by 5~$\sigma$ level. This will provide firm evidence for the past ($\sim10^5$~years ago) activity of Sgr A$^*$. Furthermore from the $nt$ value, we can trace the time history of the past flare.

\subsection{Sgr B, C, D, and E}\label{sec:SgrBCDE}
Sgr B and C are typical XRNe \citep{Ko07SgrB, Na09}. Sgr D and E are also likely to be XRNe (Nobukawa et al.2014, private communication). 
Unlike Radio Arc, contaminations of the GCXE in these regions are small, 
and hence these sources are the best targets to investigate photo-ionized astrophysical plasma and XRNe scenario. 
Since short-term time-variability was observed with the {\it Suzaku} satellite, we may add further data for the light curve of Sgr A$^*$.
However we need long time exposure because the 6.4~keV line fluxes are far smaller than that of Radio Arc.
In a long time exposure, we can see detailed line physics like Compton shoulder, Doppler broadening etc. as in the case of Radio Arc. 
The measurements of the center energy of the 6.4~keV lines may add the Galactic rotation curve data up to $\geq 100$~pc (see section~\ref{sec:GCDynamics}).

\subsection{GC Dynamics} \label{sec:GCDynamics}
\begin{figure}[btp] 
\begin{center}
\includegraphics[width=.48\linewidth]{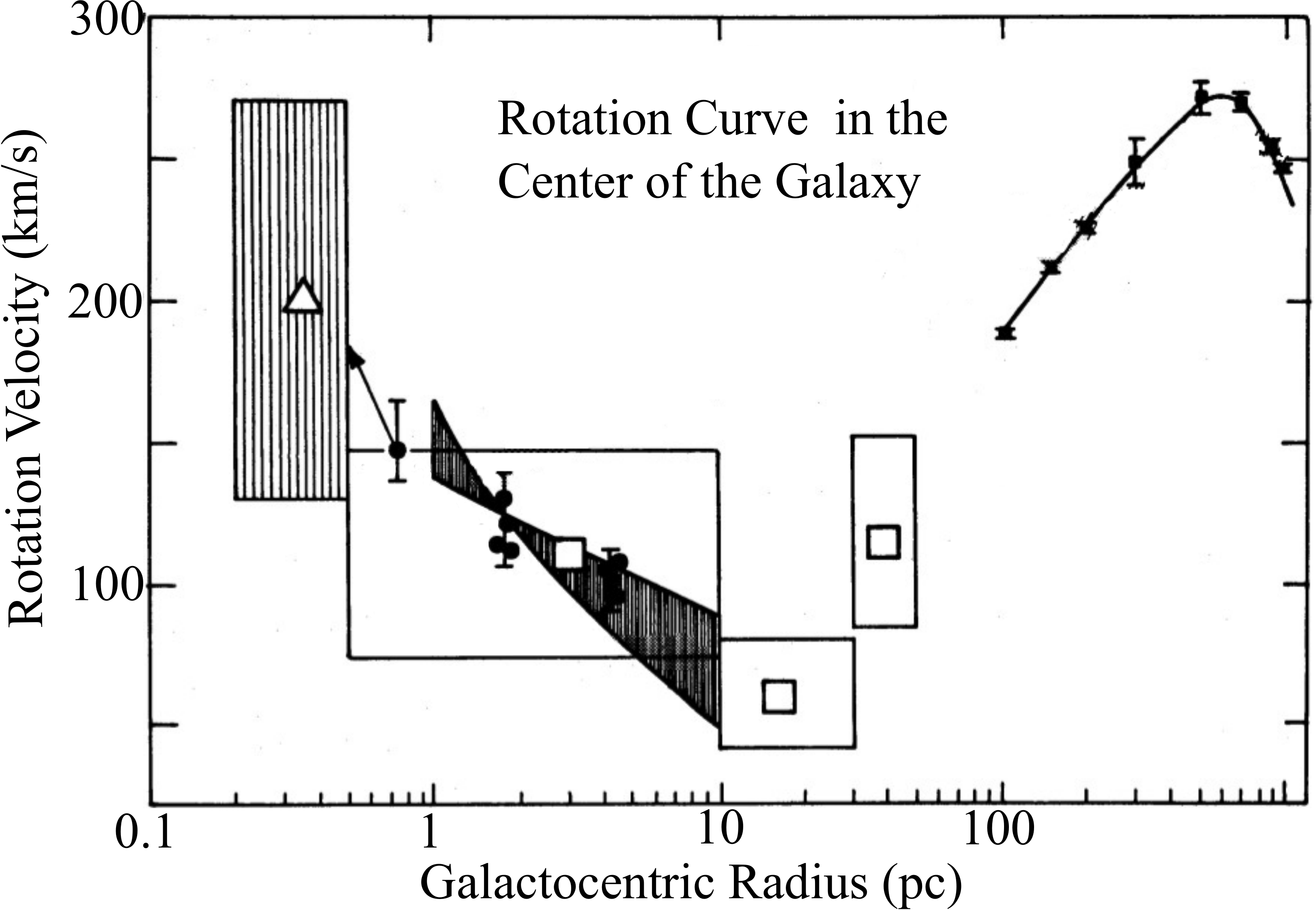}
\includegraphics[width=.46\linewidth]{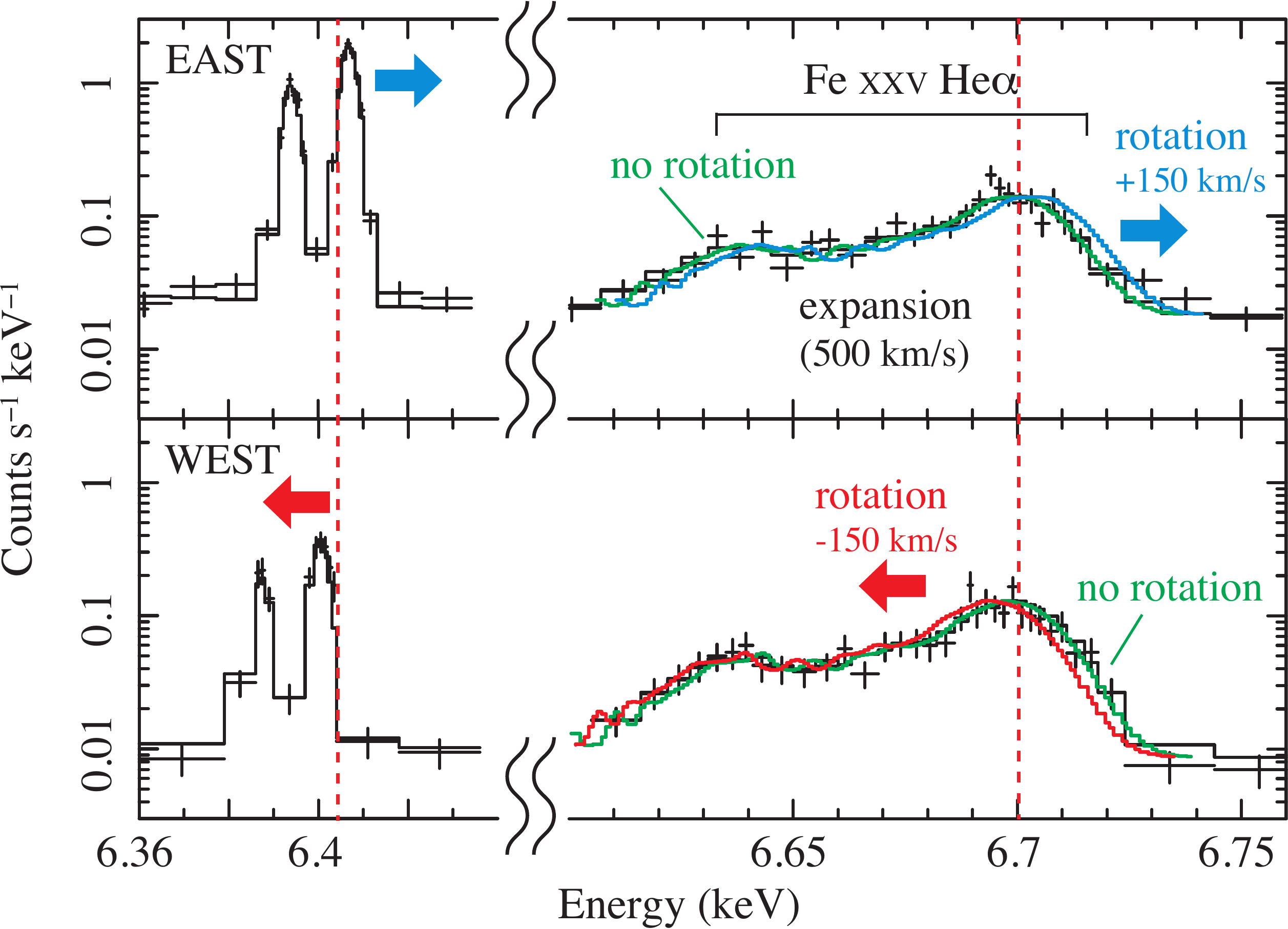}
\end{center}
\caption{
Left: Galactic rotation curve obtained with the Radio and IR observations \citep{Lu86, Cr85}. 
Right: The Doppler shifted spectra of the 6.4~keV line by the Galactic rotation at 
$l=0^{\circ}.2$ with rotation of $+150$~km~s$^{-1}$ (Upper) and 
at $l=-0^{\circ}.2$ with rotation of $-150$~km~s$^{-1}$ (Lower).
The 6.7~keV line is assumed to be expanding by 500~km~s$^{-1}$ without the Galactic rotation. 
If the line is also rotating with the 6.4~keV line, the spectra are given by the blue and red lines.} 
\label{fig:GCDynamics}
\end{figure}

With the SXS observations of the selected regions given in section \ref{sec:RadioArc}--\ref{sec:SgrBCDE}, we can obtain a systematic trend of the center  energy of the Fe~I~K$\alpha$ line as a function of the Galactic coordinate. 
The accuracy of  center energy  is $\leq$1~eV, corresponding to the velocity shift of  $\leq 50$~km~s$^{-1}$.  Thus we can measure a rotation curve of the XRNe with reasonable accuracy.

The comparison of the XRNe rotation curve with those of molecular emission line observations in radio may help to judge physical correlations between respective XRNe 
and MCs (figure~\ref{fig:GCDynamics}). 
There are not only XRNe but also no X-ray emitting molecular clouds.
Based on the accurate XRNe map including such dark clouds, we may establish the past light curve of Sgr A$^*$; which MCs in which distance are irradiated by past flare (XRNe). 

If the majority of the GCXE is diffuse plasma origin, the rotation curve of the GC hot plasma may not be necessarily the same as that of the radio band, because the hot plasma is not bounded by the gravitational potential of the GC region. The rotation curve for the GCXE hot plasma would be obtained by the accurate SXS measurement of the line centers of  Fe~XXVI~Ly$\alpha$ and fine structure lines of Fe~XXV~He$\alpha$.  The other dynamics such as expansion velocity as a function of the Galactic position will be obtained by the width measurement of these lines.
Then we examine the key question; Is it out-flowing, in-flowing or rotating differently from the CMZ ?  This may be the first information to provide the dynamics of the GC hot plasma.

\subsection{Search for Dark Matter}\label{sec:Neutrino}
Sterile neutrinos are one of the candidates for dark matter, with their mass lying in the range of 1--30~keV. With pair decay, a photon at half the energy of its mass will be observed. The Galactic center region may have high dark matter surface density and hence provides an attractive environment to search for radiative decay signals. This science will be also proposed from the  galaxy clusters  \citep {AHCluster}.
As noted in section~\ref{sec:Seren}, with {\it Suzaku}, possible sterile neutrino signal is found 
at 8.7~keV \citep{Pr10, Ch11} in the Galactic center. The scale height of the GCXE is $\sim0^{\circ}.3$--$0^{\circ}.4$ \citep{Uc13} which is far smaller than that of the dark matter distribution. Thus more suitable area  for the sterile neutrino search would be the Galactic Bulge (GB) region. In fact, faint signals at 9.4 and 10.1~keV are also found in the GB spectra (figure~\ref{fig:Neutrino}).
{\it ASTRO-H} will make many pointing observations of SXS near at the GC region (section~\ref{sec:RadioArc}--\ref{sec:SgrBCDE}). 
As by-products, we may obtain high statistics data of SXI and  HXI with the larger collecting area and wider energy range (up to $\sim 80$~keV) than those of {\it Suzaku}.
Adding all these spectra in the GC/GB regions, we may obtain unprecedented spectra, from  which possible candidate for the sterile neutrino signals may be found. 
The SXS will of course produce an unprecedented high resolution spectrum.  
Our objective is not limited to only the sterile neutrino signal, but also other unidentified line-like features, a hint of new science.


\clearpage
\begin{multicols}{2}
\end{multicols}

\end{document}